%% file: ParametricMediation.tex
\documentclass{article}

\usepackage{authblk}
\usepackage{graphicx}
\usepackage{siunitx}
\usepackage{natbib}
\usepackage[utf8x]{inputenc}
\usepackage{amsmath}
\usepackage{amssymb}
\usepackage{bm}
\usepackage{subfig}
\usepackage{pgfplots}
\usepackage{tabularx}
\usepackage{multirow}
\usepackage{booktabs}
\usepackage{array}
\usepackage{fancybox}
\usepackage{xcolor}
\usepackage{pifont}
\usepackage{picture}
\usepackage{url}
\usepackage[plain,noend]{algorithm2e}
\usepackage{tikz}
\usetikzlibrary{arrows,positioning,shapes.geometric,decorations.markings}
\tikzstyle{squarenode}=[rectangle,draw]
\tikzstyle{littlenode}=[circle,draw,minimum size=0.8cm,font=\small]
\tikzstyle{bigellipse}=[ellipse,draw,x radius=10cm, y radius=5cm]
\tikzset{negate/.style={
            decoration={markings,
            mark= at position 0.30 with {
                  \node[yshift=13pt,transform shape] (tempnode) {$\Bigg\Vert$};
                  }
              },
              postaction={decorate}
}
}

\newcommand{\ind}{\perp\!\!\!\perp}

\newcommand{\tu}[1]{{\textup{#1}}}

\begin{document}

\title{Exact parametric causal mediation analysis for a binary outcome with a binary mediator}

\author[1]{Marco Doretti}
\author[2]{Martina Raggi}
\author[3]{Elena Stanghellini}

\affil[1]{University of Perugia, Department of Political Science}
\affil[2]{University of Neuch\^{a}tel, Faculty of Economics and Business}
\affil[3]{University of Perugia, Department of Economics}

\date{}

\maketitle

\begin{abstract} 
With reference to causal mediation analysis, a parametric expression for natural direct and indirect effects is derived for the setting of a binary outcome with a binary mediator, both modelled via a logistic regression. The proposed effect decomposition operates on the odds ratio scale and does not require the outcome to be rare. It generalizes the existing ones, allowing for interactions between both the exposure and the mediator and the confounding covariates. The derived parametric formulae are flexible, in that they readily adapt to the two different natural effect decompositions defined in the mediation literature. In parallel with results derived under the rare outcome assumption, they also outline the relationship between the causal effects and the correspondent pathway-specific logistic regression parameters, isolating the controlled direct effect in the natural direct effect expressions. Formulae for standard errors, obtained via the delta method, are also given. An empirical application to data coming from a microfinance experiment performed in Bosnia and Herzegovina is illustrated.
\end{abstract}

\input{intro}

\input{decomp}

\input{data}

\input{concl}

\section*{Conflict of interest}
None

\bibliographystyle{spbasic}      
\bibliography{BiblioPM2}

\appendix

\input{appendix}

\end{document}

%% file: intro.tex
\section{Introduction}\label{sec:intro}
Mediation analysis has the general purpose to understand to what extent the overall effect of a treatment/exposure ($X$) on an outcome ($Y$) is due to the presence of an intermediate variable ($W$), called mediator, influenced by the treatment and affecting the outcome in turn. Mediation analysis has been developed both for associational~\citep{judd1981process,baron1986moderator} and causal frameworks~\citep{robins1992identifiability,Pearl2001}; see also~\cite{Geneletti2007} for an alternative approach without counterfactuals.

In a non causal framework, the first notable contribution to mediation for the linear case is due to~\cite{Cochran1938}. As is well-known, Cochran's formula decomposes the total effect of $X$ on $Y$ into the sum of products of pathway specific regression parameters, thereby opening the way to path analysis~\citep{BollenBook}. Unfortunately, this one-to-one mapping between effects and regression coefficients is lost when even minimal deviations from linearity, like the presence of an $XW$ interaction term in the model for $Y$, are introduced. Nevertheless, a number of decompositions have been proposed where the total effect can still be written as a sum of terms with a clear interpretation; see~\cite{Cox2007} for quantile regression and~\cite{StangheDoro2018} and~\cite{Lupparelli2018} for logistic and log-linear regression.

In a causal framework, formal definitions of total, direct and indirect effects were first introduced by~\cite{robins1992identifiability} and~\cite{Pearl2001}; see also~\cite{PearlCausalityBook2009} for a comprehensive overview. In these approaches, effect decompositions typically work on an additive scale. Specifically,~\cite{robins1992identifiability} introduced the so-called \emph{pure} and \emph{total} direct and indirect effects, so that two different decompositions are obtained overall. In detail, the pure direct effect and the total indirect effect sum to the total effect, and so do the total direct effect and the pure indirect effect. All these direct and indirect effects are referred to as \emph{natural} effects, in addition to the \emph{controlled} direct effect as defined in~\cite{Pearl2001}. Under certain assumptions, natural effects can be identified and estimated from observational data, even in a non-parametric setting, with a class of methods known as the mediation formula~\citep{Pearl2010med}.

For binary outcomes, ratio scales have also been recently investigated in the literature. In particular,~\cite{VdWVan2010} and~\cite{ValeriVdW2013} have defined causal effects on the odds ratio scale in a way such that the relationship between the total effect and the natural direct and indirect effects is no longer additive but multiplicative. In this framework, the aforementioned twofold decomposition is maintained and the mediation formula can still be applied, leading to parametric expressions for natural effects. Like in the associational case, such formulations lack an one-to-one mapping with regression coefficients due to non-linearity. Nevertheless, they remain appealing since they highlight the role of path-specific coefficients in a rather intuitive way, allowing, as a by-product, to isolate the controlled direct effect as a component of the natural direct effect.

It is important to notice that~\cite{VdWVan2010} and~\cite{ValeriVdW2013} base their parametric identification of natural effect odds ratios on the assumption that the outcome is rare within all the strata formed by $X$ and $W$, i.e., that $P(Y=1\mid X=x,W=w)$ is small for every $(x,w)$ configuration; see a detailed discussion in~\cite{Samoilenkoetal2018,SamoilenkoLefebvre2018rare,VdWetal2018rare}. This allows them to use the logarithmic function in place of the logistic function, corresponding in practice to approximate effects on the odds ratio scale to effects on the risk ratio scale. When the outcome is not (conditionally) rare, such an approximation is no longer valid, thereby representing a serious limitation in many empirical data analyses.

In this paper, we focus on a setting with a binary outcome and a binary mediator, both modelled via a logistic regression. We provide a novel parametric expression for the natural direct and indirect effects, on the odds ratio scale, that does not rely on the rare outcome assumption. Like the approaches developed under the rare outcome assumption, our formulation allows to appreciate the role pathway-specific coefficients play in natural effects, isolating the controlled direct effect in the natural direct effect formulae. This is an advantage compared to other exact formulations already introduced in the literature, where parametric expressions are plugged in place of the probabilities appearing in the mediation formula~\citep{Gaynoretal2018,Samoilenkoetal2018}. Furthermore, our approach is rather flexible since a single parametric object, that we name $A$-term, governs all the causal contrasts needed to identify the four natural effects.

In this framework, it is possible to account for the possible presence of any kind of confounders, that is, exposure-outcome, mediator-outcome or exposure-mediator confounders. Therefore, the proposed formulae can be used to estimate causal effects in the presence of observational data, provided that all relevant confounders are measured without error. Furthermore, they handle every possible interaction in regression models, including those between the exposure (as well as the mediator) and the confounding covariates. These interactions were not previously considered by either exact or approximate approaches; see~\cite{VdWVan2010} and~\cite{ValeriVdW2013} for a related discussion. We also derive compact formulae to compute, via the delta method, the approximate standard errors of the exact natural effect estimators. The availability of these formulae is likely to boost the use of exact estimators in place of approximate estimators in applications.

The paper is structured as follows. In Section~\ref{sec:dec}, we outline the general theory leading to the natural effect decompositions, reporting the parametric formulae for the natural direct and indirect effects on the odds ratio scale. The relationship between these effects and the correspondent pathway-specific regression coefficients is studied in Section~\ref{subsec:pathspecific}. In Section~\ref{sec:data}, we discuss an application to data gathered from a randomized microcredit experiment performed in Bosnia and Herzegovina~\citep{augsburg2015impacts}, where a plausible mediation scheme arises which was not considered in previous analyses. Finally, in Section~\ref{sec:concl} some concluding remarks are offered.

%% file: decomp.tex
\section{Parametric effect decomposition}\label{sec:dec}
\subsection{Notation and assumptions}\label{subsec:notation}
We denote the binary outcome by $Y$, the binary mediator by $W$ and the treatment/exposure, which can be of any nature, by $X$. We take a potential outcome approach~\citep{Rubin1974} and let $Y_{x}$ and $W_{x}$ be, respectively, the random variables representing the outcome and the mediator had the exposure been set, possibly contrary to the fact, to level $x$. Further, $Y_{xw}$ indicates the value of the outcome if $X$ had been set to $x$ and $W$ to $w$.

In line with the classical causal mediation framework, we make the standard assumptions needed to identify causal direct and indirect effects. Among these there are the so-called consistency and composition assumptions. Consistency states that, in the subgroup of units with $X=x$, the observed variables $Y$ and $W$ equal the potential outcome variables $Y_x$ and $W_x$ respectively~\citep{VdW2009}. In the mediation framework, consistency also requires that, for units with $X=x$ and $W=w$, $Y$ is equal to the potential outcome $Y_{xw}$~\citep{VdWVan2009}. On the other hand, composition requires that $Y_x=Y_{xW_x}$, i.e., that the potential outcome associated to the intervention $X=x$ be equal to the potential outcome associated to setting $X$ to $x$ and the mediator to $W_x$, which is the value it would have naturally attained under $X=x$~\citep{VdWVan2009}.

\begin{figure}[tb]
\begin{center}
\begin{tikzpicture}[scale=0.4,auto,->,>=stealth',shorten >=1pt,node distance=2cm] 
\node (X) {$X$};
\node (Y) [right of=X] {$Y$};
\node (W) [above right of=X,xshift=-0.4cm,yshift=-0.2cm] {$W$};
\node (T) [below right of=X,xshift=-0.4cm,yshift=+0.2cm] {$T$};
\node (V) [above of=X] {$V$};
\node (S) [above of=Y] {$S$};
\draw[->] (W) -- (Y); \draw[->] (X) -- (Y); \draw[->] (X) -- (W);
\draw[->] (T) -- (X); \draw[->] (T) -- (Y); \draw[->] (V) -- (X); \draw[->] (V) -- (W); \draw[->] (S) -- (Y); \draw[->] (S) -- (W);
\end{tikzpicture}
\end{center}
\caption{Causal mediation setting with exposure-outcome ($T$), mediator-outcome ($S$) and exposure-mediator ($V$) confounders\label{fig:medconf}}
\end{figure}
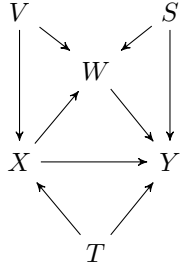

A number of assumptions concerning confounding are also required which are graphically summarized in Figure~\ref{fig:medconf}. Specifically, we assume a set of covariates $C=(T,S,V)$ suffices to remove the exposure-outcome ($T$), the mediator-outcome ($S$) and the exposure-mediator ($V$) confounding. In the language of conditional independence~\citep{Dawid1979}, this corresponds to the conditional independence statements

i) $Y_{xw}\ind X \mid T$,

ii) $Y_{xw} \ind W \mid (X,S)$,

iii) $W_{x} \ind X \mid V$,

\noindent which have to hold for every level $x$ and $w$. Another necessary assumption, sometimes termed cross-world independence~\citep{Steen2018}, is encoded by the conditional independence statement 

iv) $Y_{xw} \ind W_{x^{\star}} \mid C$,

\noindent for all $x$, $x^{\star}$ and $w$. This assumption means in practice that none of the variables tackling the mediator-outcome confounding can be affected by the treatment (which would correspond to an arrow from $X$ to $S$ in Figure~\ref{fig:medconf}). Such an effect would compromize the identification of natural effects~\citep{ValeriVdW2013,VdWVan2010}, unless alternative methods to handle variables that are simultaneously confounders and mediators are introduced~\citep{Robins1986,daniel2015causal,Steenetal2017multi}. In what follows, to simplify formulae we will use the condensed notation $Z=(T,S)$. Without loss of generality, we can think of $Z$ and $V$ as of univariate random variables. This is equivalent to assume that one covariate addresses both the exposure-outcome and the mediator-outcome confounding and another covariate manages the exposure-mediator confounding. Results for the case of multiple $Z$ and $V$ follow in a straightforward way; see Appendix~\ref{app:math}.

In this framework, with reference to a situation where the exposure is changed from a reference level $x^{\star}$ to another level $x$,~\cite{VdWVan2010} introduced the definitions of causal effects on the odds ratio scale. Specifically, the controlled direct effect is defined as
\begin{equation}\label{eq:cdedef}
\tu{OR}_{x,x^{\star}\mid c}^{\tu{CDE}}(w) = \frac{P(Y_{xw}=1\mid c)/P(Y_{xw}=0\mid c)}{P(Y_{x^{\star}w}=1\mid c)/P(Y_{x^{\star}w}=0\mid c)}
\end{equation}
and describes the causal effect of the exposure on the outcome while the mediator is kept to level $w$ for every unit~\citep{VdWVan2010}. In contrast, the pure natural direct effect
\begin{equation}\label{eq:ndedef}
\tu{OR}_{x,x^{\star}\mid c}^{\tu{PNDE}} = \frac{P(Y_{xW_{x^{\star}}}=1\mid C=c)/P(Y_{xW_{x^{\star}}}=0\mid C=c)}{P(Y_{x^{\star}W_{x^{\star}}}=1\mid C=c)/P(Y_{x^{\star}W_{x^{\star}}}=0\mid C=c)}
\end{equation}
quantifies the same effect when keeping the mediator to $W_{x^{\star}}$, that is, to the level it would have naturally attained for each unit under the exposure level $x^{\star}$~\citep{VdWVan2010}. The total natural indirect effect is given by
\begin{equation}\label{eq:niedef}
\tu{OR}_{x,x^{\star}\mid c}^{\tu{TNIE}} = \frac{P(Y_{xW_{x}}=1\mid C=c)/P(Y_{xW_{x}}=0\mid C=c)}{P(Y_{xW_{x^{\star}}}=1\mid C=c)/P(Y_{xW_{x^{\star}}}=0\mid C=c)}
\end{equation}
and compares the odds of $Y=1$ had the exposure been set to $x$ and the mediator been set to the value that it would  have naturally taken if the exposure had been set to $x$ ($W_x$) against the odds of $Y=1$ had the exposure been set to $x$ but the mediator been set to the value that it would  have naturally achieved if exposure had been set to $x^{\star}$ ($W_{x^{\star}}$). In other terms, the total natural indirect effect measures the causal effect on the outcome of moving the mediator from $W_{x^{\star}}$ to $W_x$ while keeping the exposure $X$ fixed to level $x$~\citep{VdWVan2010}.

With regard to the other decomposition, the total natural direct effect can be defined as
\begin{equation}\label{eq:tndedef}
 \tu{OR}_{x,x^{\star}\mid c}^{\tu{TNDE}} = \frac{P(Y_{xW_{x}}=1\mid C=c)/P(Y_{xW_{x}}=0\mid C=c)}{P(Y_{x^{\star}W_{x}}=1\mid C=c)/P(Y_{x^{\star}W_{x}}=0\mid C=c)},
\end{equation}
whereas the pure natural indirect effect is
\begin{equation}\label{eq:pniedef}
 \tu{OR}_{x,x^{\star}\mid c}^{\tu{PNIE}} = \frac{P(Y_{x^{\star}W_{x}}=1\mid C=c)/P(Y_{x^{\star}W_{x}}=0\mid C=c)}{P(Y_{x^{\star}W_{x^{\star}}}=1\mid C=c)/P(Y_{x^{\star}W_{x^{\star}}}=0\mid C=c)}.
\end{equation}
In line with~\eqref{eq:ndedef}, the former represents the causal effect on the outcome of moving the exposure from $x^{\star}$ to $x$ under an external intervention keeping the mediator to $W_x$, the level it would have naturally taken for each unit under the exposure level $x$. The latter, in analogy with~\eqref{eq:niedef}, compares the causal effect on the outcome of moving the mediator from $W_x$ to $W_{x^{\star}}$, with exposure $X$ set to level $x^{\star}$.

Finally, the total causal effect is defined as
\[
\tu{OR}_{x,x^{\star}\mid c}^{\tu{TE}} = \frac{P(Y_{x}=1\mid C=c)/P(Y_{x}=0\mid C=c)}{P(Y_{x^{\star}}=1\mid C=c)/P(Y_{x^{\star}}=0\mid C=c)}
\]
and can be decomposed in the multiplicative fashion
\[
\begin{split}
\tu{OR}_{x,x^{\star}\mid c}^{\tu{TE}}&= \tu{OR}_{x,x^{\star}\mid c}^{\tu{PNDE}} \times \tu{OR}_{x,x^{\star}\mid c}^{\tu{TNIE}} \\
&= \tu{OR}_{x,x^{\star}\mid c}^{\tu{TNDE}} \times \tu{OR}_{x,x^{\star}\mid c}^{\tu{PNIE}} \\
\end{split}
\]
or, as it is often presented, in an additive fashion on the logarithmic scale. The choice of which natural effect to use, and thus which decomposition, depends on the research questions one aims to investigate~\citep{hafeman2009opening}. Combining the assumptions given in the Subsection~\ref{subsec:notation}, the causal effects defined above can be non-parametrically identified by using the Pearl's mediation formula~\citep{Pearl2001,Pearl2010med,VdWVan2010,ValeriVdW2013}.


\subsection{Parametric formulae for causal natural effects}\label{subsec:parne}
As mentioned in the Introduction, parametric formulae can be derived which express the natural effects as a function of the coefficients of the logistic regression models for $Y$ and $W$, that can be easily estimated from observational data. We first present results for the case of absence of covariates, i.e. for $C=\emptyset$. Then, the more general case will be addressed. Specifically, we can formulate the two models as
\begin{equation}\label{eq:ynocov}
\log\frac{P(Y=1\mid X=x,W=w)}{P(Y=0\mid X=x,W=w)} = \beta_{0}+\beta_{x}x+\beta_{w}w + \beta_{xw}xw
\end{equation}
and
\begin{equation}\label{eq:wnocov}
\log\frac{P(W=1\mid X=x)}{P(W=0\mid X=x)} = \gamma_{0}+\gamma_{x}x.
\end{equation}

In this setting, the logarithm of the controlled direct effect~\eqref{eq:cdedef} can be identified by
\[
\log \tu{OR}_{x,x^{\star}}^{\tu{CDE}(w)} = ( \beta_x + \beta_{xw}w ) (x-x^{\star})
\]
without the need to invoke the rare outcome assumption~\citep{VdWVan2010}. Here, letting
\[
\begin{split}
e_y(x,w) &= \exp(\beta_0+\beta_xx+\beta_ww+\beta_{xw}xw) \\
e_w(x) &= \exp(\gamma_0+\gamma_xx)
\end{split}
\]
and
\begin{equation}\label{eq:axxsnocov}
A_{x,x^{\star}} = \frac{\exp(\beta_{w}+\beta_{xw}x)e_w(x^{\star})\{1+e_y(x,0)\}+1+e_y(x,1)}{e_w(x^{\star})\{1+e_y(x,0)\}+1+e_y(x,1)},
\end{equation}
we show that the parametric expressions for the logarithm of the causal effects defined in~\eqref{eq:ndedef} and~\eqref{eq:niedef} are given by
\begin{equation}\label{eq:ndenocov}
\log \tu{OR}_{x,x^{\star}}^{\tu{PNDE}} = \beta_{x}(x-x^{\star}) + \log\frac{A_{x,x^{\star}}}{A_{x^{\star},x^{\star}}}
\end{equation}
and
\begin{equation}\label{eq:nienocov}
\log \textup{OR}_{x,x^{\star}}^{\textup{TNIE}} = \log\frac{A_{x,x}}{A_{x,x^{\star}}}.
\end{equation}
Similarly, formulae for the logarithm of the effects in~\eqref{eq:tndedef} and~\eqref{eq:pniedef} can be written as
\begin{equation}\label{eq:tndenocov}
\log \tu{OR}_{x,x^{\star}}^{\tu{TNDE}} =\beta_x(x-x^{\star})+ \log \frac{A_{x,x}}{A_{x^{\star},x}},
\end{equation}
and
\begin{equation}\label{eq:pnienocov}
\log \tu{OR}_{x,x^{\star}}^{\tu{PNIE}} = \log \frac{A_{x^{\star},x}}{A_{x^{\star},x^{\star}}},
\end{equation}

The mathematical derivations leading to~\eqref{eq:ndenocov} and~\eqref{eq:nienocov} are given in Appendix~\ref{app:math}, with those for~\eqref{eq:tndenocov} and~\eqref{eq:pnienocov} following in a straightforward way. Combining~\eqref{eq:ndenocov} and~\eqref{eq:nienocov} (or alternatively,~\eqref{eq:tndenocov} and~\eqref{eq:pnienocov}) one immediately obtains the identification expression for the logarithm of the total casual effect, which is
\begin{equation}\label{eq:tenocov}
\log \tu{OR}_{x,x^{\star}}^{\tu{TE}} = \beta_{x}(x-x^{\star}) + \log\frac{A_{x,x}}{A_{x^{\star},x^{\star}}}.
\end{equation}

The parametric object in Equation~\eqref{eq:axxsnocov} is called the $A$-term. Notice that its first subscript refers to the value of $X$ in the model for $Y$ given $X$ and $W$, while the second one refers to the value of $X$ in the model for $W$ given $X$. The main advantage of this compact notation is the fact that it creates a clear parallel between the contrasts contained in the definitions~\eqref{eq:ndedef}-\eqref{eq:pniedef} and those contained in the identification formulae~\eqref{eq:ndenocov}-\eqref{eq:pnienocov}. In a non causal framework,~\cite{StangheDoro2018} have shown that $A_{x,x}$ is the inverse of the risk ratio of $\bar W = 1-W$ for varying $Y$ when $X=x$, that is
\begin{equation}\label{eq:invrr}
A_{x,x} = \frac{P(\bar W=1 \mid Y=0,X=x)}{P(\bar W=1 \mid Y=1,X=x)}.
\end{equation}
However, it is important to underline this interpretation holds only when the two subscripts of the $A$-term take the same value.

The first of the two parametric decompositions above generalizes the one by~\cite{ValeriVdW2013} developed under the rare outcome assumptions. Specifically, it holds that the exponential of~\eqref{eq:ndenocov} tends to the pure natural direct effect in~\citet[p.150]{ValeriVdW2013} if $P(Y=1\mid X=x,W=w)$ (or, equivalently, the exponentiated linear predictor $e_y(x,w)$) tends to zero for every $(x,w)$ configuration. A similar argument applies to the total natural indirect effect, though with some distinctions regarding the set of configurations tending to zero. A formal account about these limits concerning also the other decomposition is contained in Appendix~\ref{app:limits}. Furthermore, like in~\cite{ValeriVdW2013}, our formulation expresses the odds ratio natural direct effects (pure and total) as the product between the controlled direct effect $\tu{OR}_{x,x^{\star}\mid c}^{\tu{CDE}}(0)$ and a residual term. 

The extension of the above setting to the parametric inclusion of covariates $C$ is immediate. In detail, if Equations~\eqref{eq:ynocov} and~\eqref{eq:wnocov} are modified to account for these additional covariates and for all their possible interactions, i.e., to
\begin{equation}\label{eq:y}
\begin{split}
\log\frac{P(Y=1\mid X=x,W=w,Z=z)}{P(Y=0\mid X=x,W=w,Z=z)} &= \beta_{0}+\beta_{x}x+\beta_{w}w+\beta_zz+\beta_{xw}xw \\
												&+\beta_{xz}xz+\beta_{wz}wz+\beta_{xwz}xwz
\end{split}
\end{equation}
and
\begin{equation}\label{eq:w}
\log\frac{P(W=1\mid X=x,V=v)}{P(W=0\mid X=x,V=v)} = \gamma_{0}+\gamma_{x}x+\gamma_vv+\gamma_{xv}xv,
\end{equation}
then also the natural effects become conditional on the covariate configuration $C=c$. Specifically, we have
\begin{equation}\label{eq:ndecov}
\log \tu{OR}_{x,x^{\star}\mid c}^{\tu{PNDE}} = (\beta_x+\beta_{xz}z)(x-x^{\star}) + \log\frac{A_{x,x^{\star}\mid c}}{A_{x^{\star},x^{\star}\mid c}}
\end{equation}
and
\begin{equation}\label{eq:niecov}
\log \tu{OR}_{x,x^{\star}\mid c}^{\tu{TNIE}} = \log\frac{A_{x,x\mid c}}{A_{x,x^{\star}\mid c}},
\end{equation}
where, denoting the exponentiated linear predictors by the compact forms
\[
\begin{split}
e_y(x,w,z) &= \exp(\beta_0+\beta_xx+\beta_ww+\beta_zz+\beta_{xw}xw+\beta_{xz}xz+\beta_{wz}wz+\beta_{xwz}xwz) \\
e_w(x,v) &= \exp(\gamma_0+\gamma_xx+\gamma_vv+\gamma_{xv}xv),
\end{split}
\]
the conditional version of~\eqref{eq:axxsnocov} is given by
\begin{equation}\label{eq:acov}
A_{x,x^{\star}\mid c} = \frac{\exp(\beta_w+\beta_{xw}x+\beta_{wz}z+\beta_{xwz}xz)e_w(x^{\star},v)\{1+e_y(x,0,z)\}+1+e_y(x,1,z)}{e_w(x^{\star},v)\{1+e_y(x,0,z)\}+1+e_y(x,1,z)}.
\end{equation}
The proof follows immediately from the previous one and is also reported in Appendix~\ref{app:math}. Similarly to the previous case, the parametric identification of the other causal estimands is obtained from the conditional versions of Equations~\eqref{eq:tndenocov} and~\eqref{eq:pnienocov}. 

Clearly, the consistent estimation of these causal effects can be achieved by simply plugging-in the parameter estimates obtained from the fitted logistic regression models in the formulae above. The variance-covariance matrix of the effect estimators can be obtained via the delta method~\citep{Oehlert1992}. The explicit formulae for the first-order delta approximation of such a matrix are reported in Appendix~\ref{app:delta}.


\subsection{Links with the pathway-specific coefficients}\label{subsec:pathspecific}
The parametric formulae given in Section~\ref{subsec:parne} link natural effects to their pathway-specific regression coefficients in an explicit way. Indeed, although the general expression of the $A$-term involves the whole set of parameters, it is possible to spot a particular behavior of the natural effects when the three pathway-specific sets of parameters are null, that is, when: (i) $\beta_x=0=\beta_{xw}$, (ii) $\beta_w=0=\beta_{xw}$, and (iii) $\gamma_x=0$.

In case (i), it can be noticed that $\log \tu{OR}_{x,x^{\star}}^{\tu{PNDE}}$ and $\log \tu{OR}_{x,x^{\star}}^{\tu{TNDE}}$ are null, and so is $\log \tu{OR}_{x,x^{\star}}^{\tu{CDE(0)}}$. In summary, for this case
\[
\log \tu{OR}_{x,x^{\star}}^{\tu{TE}}=\log \tu{OR}_{x,x^{\star}}^{\tu{TNIE}} =\log \tu{OR}_{x,x^{\star}}^{\tu{PNIE}}.
\]

In case (ii), the $A$-term is equal to 1 for all the combinations of its subscripts, so that every pair of $A$-terms generates a null contrast on the logarithmic scale. As a consequence, we have that $\log \tu{OR}_{x,x^{\star}}^{\tu{TNIE}}=0=\log \tu{OR}_{x,x^{\star}}^{\tu{PNIE}}$ and that
\[
\log \tu{OR}_{x,x^{\star}}^{\tu{TE}}=\log \tu{OR}_{x,x^{\star}}^{\tu{CDE(0)}}=\log \tu{OR}_{x,x^{\star}}^{\tu{PNDE}}=\log \tu{OR}_{x,x^{\star}}^{\tu{TNDE}}=\beta_x(x-x^{\star}).
\]

In case (iii), again the two natural indirect effects vanish simultaneously, so that
\[
\log \tu{OR}_{x,x^{\star}}^{\tu{TE}}=\log \tu{OR}_{x,x^{\star}}^{\tu{PNDE}}=\log \tu{OR}_{x,x^{\star}}^{\tu{TNDE}}.
\]
However, these quantities are not necessarily equal to $\log \tu{OR}_{x,x^{\star}}^{\tu{CDE(0)}}=\beta_x(x-x^{\star})$ like in the previous case. 

It is useful to notice that such a clear correspondence between model parameters and direct/indirect effects does not hold with associational (non-causal) effects. In particular, for non-linear models it is known that the condition $\gamma_{x}=0$ does not guarantee the marginal and conditional effects of $X$ on $Y$ be equal. For logistic regression, this is related to the well-known fact that odds ratios are non-collapsible association measures~(\citealt{Greenlandetal1999}). Also, notice that the condition $\beta_{xw}=0$ does not generally imply that $\log \tu{OR}_{x,x^{\star}}^{\tu{PNDE}}$ equals $\log \tu{OR}_{x,x^{\star}}^{\tu{TNDE}}$ or that $\log \tu{OR}_{x,x^{\star}}^{\tu{PNIE}}$ equals $\log \tu{OR}_{x,x^{\star}}^{\tu{TNIE}}$. These implications hold when a linear model for the outcome is specified. For binary outcomes, the rare outcome assumption or, alternatively, the specification of a log-linear regression model is required. For a more detailed dissertation within the four-way decomposition framework see~\cite{VdW2014}.

Another interesting situation arises when $x^{\star}=0$, including (but not limited to) the case of binary exposures. In this case, we can notice that $\log \tu{OR}_{x,0}^{\tu{CDE(0)}}$ is obtained evaluating $\log \tu{OR}_{x,0}^{\tu{TE}}$ in $\beta_w=0=\beta_{xw}$. Furthermore, $\log \tu{OR}_{x,0}^{\tu{PNIE}}$ is obtained from $\log \tu{OR}_{x,0}^{\tu{TE}}$ evaluated in $\beta_x=0=\beta_{xw}$, whereas $\log \tu{OR}_{x,0}^{\tu{PNDE}}$ equals $\log \tu{OR}_{x,0}^{\tu{TE}}$ evaluated in $\gamma_x=0$.

%% file: data.tex
\section{Application to Bosnian microcredit data}\label{sec:data}
Microcredit, as the main tool of microfinance, makes credit accessible to those individuals, often termed ``unbankables'', that are considered too risky and financially unreliable to access regular loans granted by financial institutions. The main purpose of microcredit is the enhancement of the overall ``bankability''  - that is, the capability to attract loans from banks or other microfinance institutions (MFIs) - of the financially disadvantaged individuals. We here offer an empirical application of the derived analytical results to a microcredit experiment implemented in Bosnia and Herzegovina by~\cite{augsburg2015impacts}; see also~\cite{banerjee2015six} for details about the more general project involving similar experiments in other countries. This study was performed during the period 2009-2010 and was addressed to a particular segment of unbankable people formed by the potential clients of a well-established MFI of the country. 

The main goal of the experiment was to evaluate the impact of randomly allocated microcredit loans not only on clients' bankability but also on a number of other socioeconomic outcomes including self-employment, business ownership, income, time worked, consumption and savings. At baseline, clients were selected to take part to the experiment and enrolled in a pre-intervention survey in order to collect main information concerning them and their household. Then, they were randomly assigned to the exposure (access to the microloan) or control group. After 14 months, the research team conducted a follow-up survey on the same respondents recruited at baseline. In total, 995 respondents were interviewed at the two waves. In the description of the experiment contained in~\cite{augsburg2015impacts}, non-compliance issues appear to be absent, but a 17\% loss to follow-up was registered. However, the reasons leading to client dropout, together with formal analyses reported in the paper, do not induce to think that a non-ignorable missigness mechanism~\citep{Molenberghsetal2008} occurred. The average microloan amount was equal to 1'653 Bosnian marks (BAM, with an exchange rate at baseline of US\$1 to BAM 1.634) with an average maturity of 57 weeks.

Since individuals were free to use money from the loan for business activities as well as for household consumption, a positive effect of the financing policy on many of the above-mentioned socioeconomic indicators was found~\citep{augsburg2015impacts}. However, some of these measures can be reasonably thought of not only as final outcomes, but also as determinants of clients' future credit attractiveness, lending themselves to the role of possible mediators of the overall effect of microcredit on bankability. In particular,~\cite{banerjee2015six} acknowledge, though without any formal analysis, business ownership as the main candidate as a mediator variable. Indeed, a business can be started or maintained thanks to the initial microcredit financing, and business owners are usually more likely to access the loan market than others. In line with the hypothesis above which involves a binary mediator, we here compute the two odds ratio decompositions of the causal total effect into the natural effects.

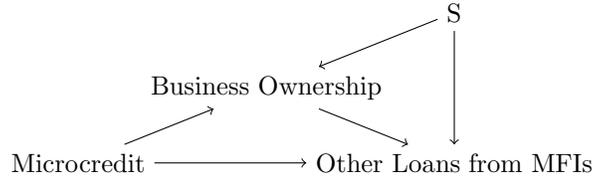
\begin{figure}[tb]
\begin{center}
 \begin{tikzpicture}
    \node (A) at (0,0) {Microcredit};
    \node (B) at (2.5,1) {Business Ownership};
    \node (C) at (5,0) {Other Loans from MFIs};
    \node (D) at (5,2) {S};

    \path [->] (A) edge node[above] {} (B);
    \path [->] (B) edge node[above] {} (C);
    \path [->] (A) edge node[below] {} (C);
    \path [->] (D) edge node[left]  {} (C);
    \path [->] (D) edge node[left]  {} (B);
\end{tikzpicture}
\end{center}
\caption{Causal mediation setting for the Bosnian microcredit study}\label{fig:data}
\end{figure}

We make use of the notation of Section~\ref{sec:dec} and denote by $X$ the binary exposure taking value 1 if the client gets the microcredit financing at baseline, by $Y$ the binary outcome taking value 1 for clients who have access to at least one new credit line at follow-up, and by $W$ the binary mediator with value 1 for units owning a personal business, which is also measured at follow-up. Notice that individuals with $Y=1$ might have received loans from an MFI as well as from other traditional institutions like banks. A graphical representation of the setting under investigation is shown in Figure \ref{fig:data}. 

From Figure~\ref{fig:data}, it is possible to note that no exposure-outcome and exposure-mediator confounders are included in the analysis since the exposure assignment is randomized. On the contrary, the set of possible mediator-outcome confounders $S$ needs to be taken into account. Some preliminary research combined with subject matter considerations led to include in $S$ client's age ($A$), educational level ($U$) and number of active loans $(L)$. In particular, age is measured in years while educational level is coded as a binary variable taking value 1 for individuals with at least a university degree. All these covariates are measured at baseline and can be considered as pre-treatment variables. This should ensure that the cross-world independence assumption (see Section~\ref{sec:dec}) holds, since none of the variables in $S$ is causally affected by the exposure~\citep{VdWVan2009,Steen2018}. In the sample, age ranges from 17 to 70 years, with an average of 37.81 years and a median of 37 years. The first and third quartiles are 28 and 47 years respectively. Further, only 5\% of sample units own a university degree, while only 4\% have three or more active loans at baseline. All the sample marginal probabilities for $X$, $W$ and $Y$ are close to 0.5. In detail, we have $P(X=1)=0.55$, $P(W=1)=0.54$ and $P(Y=1)=0.57$. Furthermore, we have $P(Y=1 \mid X=0,W=0)=0.24$, $P(Y=1 \mid X=1,W=0)=0.67$, $P(Y=1 \mid X=0,W=1)=0.41$ and $P(Y=1 \mid X=1,W=1)=0.84$, in contrast with the rare outcome assumption, which is clearly violated also conditionally on the covariates $S$.

Table~\ref{tab:regr} contains the output of the fitted logistic regression models for the outcome and the mediator, whereas Table~\ref{tab:scen2} shows the estimates, together with their variability measures, of the causal effects obtained from these model parameters. The effects refer to individuals with median age and all the most common patterns of the other covariates. The asymptotic standard errors and confidence intervals are constructed using the delta method as illustrated in Appendix~\ref{app:delta}. As in standard analyses on odds ratios, the 95\% confidence intervals are first built on the logarithmic scale and then exponentiated. Also, the p-values refer to tests where the null hypotheses are formulated on the logarithmic scale, that is, that log-odds ratios are equal to zero. In the outcome model, the presence of interaction terms involving the confounders was explored, but none of these effects resulted statistically significant or, to the best of our judgment, worth to be added to the model. 

\begin{table}[tb]
\centering
\begin{tabular}{lccccc}
\toprule
 & \multicolumn{5}{c}{$Y\sim\beta_{0}+\beta_{x}X+\beta_{w}W+\beta_{xw}XW+\beta_{a}A+\beta_{u}U+\beta_{l}L$} \\
 \cmidrule(l{3pt}r{3pt}){2-6}
 		& Estimate 	& Std. Error & \multicolumn{2}{c}{95\% Conf. Interval} 	& p-value \\
 \cmidrule(l{3pt}r{3pt}){2-6}
$\beta_{0}$	& -1.542	& 0.290 	&-2.118	& -0.981  		& 0.000 \\
$\beta_{x}$	&  1.903	& 0.213  	&1.492 	& 2.327   		& 0.000 \\ 
$\beta_{w}$	&  0.758 	& 0.211  	&0.349  	&1.175    		& 0.000 \\
$\beta_{xw}$	&  0.137 	& 0.296 	&-0.444  	&0.718    		& 0.643 \\
$\beta_{a}$	&  0.008 	& 0.006 	&-0.004  	&0.020    		& 0.214 \\
$\beta_{u}$	& -1.001 	& 0.363 	&-1.729 	&-0.299    		& 0.006 \\
$\beta_{l}$	&  0.185  	& 0.085  	&0.020  	&0.355    		& 0.029 \\
\midrule
& \multicolumn{5}{c}{$W\sim \gamma_{0}+\gamma_{x}X$} \\
 \cmidrule(l{3pt}r{3pt}){2-6}
		& Estimate 	& Std. Error & \multicolumn{2}{c}{95\% Conf. Interval} 	& p-value\\ 
 \cmidrule(l{3pt}r{3pt}){2-6}
$\gamma_{0}$ & 0.027  	&  0.095	& -0.159	&  0.213		& 0.776 \\
$\gamma_{x}$	& 0.262   	&  0.128	&  0.011	&  0.513		& 0.041\\
\bottomrule
\end{tabular}
\caption{Results from the fitted logistic models for the outcome and the mediator}\label{tab:regr}
\end{table}

Table~\ref{tab:regr} shows that all the estimated coefficients related to the mediation pathways $X\rightarrow W\rightarrow Y$ and $X\rightarrow Y$ are positive and statistically significant, with the exception of the interaction $\hat{\beta}_{xw}$, which is positive but not significant (p-value 0.643). However, it is possible to notice a relevant difference in the coefficient magnitudes. Indeed, $\hat{\beta}_{w}$ and $\hat{\gamma}_{x}$ are much smaller than $\hat{\beta}_{x}$, suggesting that the natural direct effect is the dominant component of the total effect. This is confirmed by the results in Table~\ref{tab:scen2} which are rather stable across the covariate patterns examined. Specifically, the estimated natural direct effects always lie between 6.647 and 6.868, whereas all the estimates of natural indirect effects range between 1.048 and 1.059. In this setting, the pure (total) natural direct effect is the effect of the microloan keeping business ownership to the level it would naturally take in the absence (presence) of the microloan itself. Similarly, the total (pure) natural indirect effect is the effect of moving business ownership from its natural level under no microloan to that under the microloan administration, while keeping the microloan fixed to be present (absent).

All the 95\% confidence intervals for the direct effects are far away from 1, corresponding to highly significant natural direct effects. On the contrary, all the 95\% confidence intervals for the indirect effects barely contain 1, corresponding to p-values around 6\% or 7\%. The low magnitude of the natural indirect effects might be due to the relatively limited temporal distance occurring between the baseline and the follow-up measurement occasions. Indeed, also from the original study by~\cite{augsburg2015impacts} it seems that a 14-month period may be not long enough to register any relevant effect of microcredit on business ownership. We have also replicated these results starting from an alternative outcome model where the $XW$ interaction is removed. In this model, the main effects modify to $\hat{\beta}_{x}=1.974$ (s.e. 0.149) and $\hat{\beta}_{w}=0.828$ (s.e. 0.149), while the other parameters do not sensibly change. The causal odds ratios resulting from this model are substantially equivalent to the previous ones.

\begin{table}[t]
\centering
\begin{tabularx}{\linewidth}{lXXXXcXXXXc}
\toprule
 & \multicolumn{5}{c}{$A=37,U=0,L=0$} & \multicolumn{5}{c}{$A=37,U=1,L=0$}\\
 \cmidrule(l{3pt}r{3pt}){2-6} \cmidrule(l{3pt}r{3pt}){7-11}
 & Est. & SE & \multicolumn{2}{c}{95\% CI} & p-value & Est. & SE & \multicolumn{2}{c}{95\% CI} & p-value\\ 
 \cmidrule(l{3pt}r{3pt}){2-6} \cmidrule(l{3pt}r{3pt}){7-11} 
$\tu{OR}_{1,0\mid c}^{\tu{PNDE}}$   & 6.652  & 0.953  &  5.024 &  8.809 & 0.000 &  6.796  &  0.988  & 5.112  &   9.036 & 0.000\\
$\tu{OR}_{1,0\mid c}^{\tu{TNDE}}$    & 6.717  & 0.965  &  5.069 &  8.901 & 0.000 &  6.868  &  1.019  & 5.134  &   9.187 & 0.000\\
$\tu{OR}_{1,0\mid c}^{\tu{PNIE}}$   & 1.049  & 0.028  &  0.996 &  1.105 & 0.072 &  1.048  &  0.027  & 0.996  &   1.102 & 0.071\\
$\tu{OR}_{1,0\mid c}^{\tu{TNIE}}$    & 1.059  & 0.033  &  0.997 &  1.125 & 0.063 &  1.059  &  0.033  & 0.997  &   1.125 & 0.063\\
$\tu{OR}_{1,0\mid c}^{\tu{TE}}$    & 7.046  & 1.023  &  5.302 &  9.364 & 0.000 &  7.197  &  1.071  & 5.376  &   9.635 & 0.000\\
\midrule
 & \multicolumn{5}{c}{$A=37,U=0,L=1$} & \multicolumn{5}{c}{$A=37,U=1,L=1$}\\
 \cmidrule(l{3pt}r{3pt}){2-6} \cmidrule(l{3pt}r{3pt}){7-11}
 & Est. & SE & \multicolumn{2}{c}{95\% CI} & p-value & Est. & SE & \multicolumn{2}{c}{95\% CI} & p-value\\ 
 \cmidrule(l{3pt}r{3pt}){2-6} \cmidrule(l{3pt}r{3pt}){7-11}
$\tu{OR}_{1,0\mid c}^{\tu{PNDE}}$  & 6.647  & 0.954 &  5.017 &  8.806 & 0.000 &  6.757  &  0.976 &   5.091 &  8.969 & 0.000\\
$\tu{OR}_{1,0\mid c}^{\tu{TNDE}}$  & 6.709  & 0.962 &  5.065 &  8.887 & 0.000 &  6.828  &  1.005 &   5.118 &  9.111 & 0.000\\
$\tu{OR}_{1,0\mid c}^{\tu{PNIE}}$   & 1.049  & 0.028 &  0.996 &  1.106 & 0.073 &  1.048  &  0.027 &   0.996 &  1.103 & 0.071\\
$\tu{OR}_{1,0\mid c}^{\tu{TNIE}}$   & 1.059  & 0.033 &  0.997 &  1.125 & 0.062 &  1.059  &  0.033 &   0.997 &  1.125 & 0.063\\
$\tu{OR}_{1,0\mid c}^{\tu{TE}}$       & 7.039  & 1.022 &  5.296 &  9.356 & 0.000 &  7.157  &  1.057 &   5.358 &  9.559 & 0.000\\
\midrule
& \multicolumn{5}{c}{$A=37,U=0,L=2$} & \multicolumn{5}{c}{$A=37,U=1,L=2$}\\
 \cmidrule(l{3pt}r{3pt}){2-6} \cmidrule(l{3pt}r{3pt}){7-11}
 & Est. & SE & \multicolumn{2}{c}{95\% CI} & p-value & Est. & SE & \multicolumn{2}{c}{95\% CI} & p-value\\ 
 \cmidrule(l{3pt}r{3pt}){2-6} \cmidrule(l{3pt}r{3pt}){7-11}
$\tu{OR}_{1,0\mid c}^{\tu{PNDE}}$   & 6.647  & 0.957 & 5.012 &  8.815& 0.000 & 6.723 &  0.967 &  5.072 &   8.913 & 0.000\\
$\tu{OR}_{1,0\mid c}^{\tu{TNDE}}$   & 6.708  & 0.962 & 5.065 &  8.885& 0.000 & 6.793 &  0.992 &  5.103 &   9.044 & 0.000\\
$\tu{OR}_{1,0\mid c}^{\tu{PNIE}}$   & 1.049  & 0.028 & 0.996 &  1.106& 0.073 & 1.048 &  0.027 &  0.996 &   1.103 & 0.071\\
$\tu{OR}_{1,0\mid c}^{\tu{TNIE}}$   & 1.059  & 0.033 & 0.997 &  1.125& 0.062 & 1.059 &  0.033 &  0.997 &   1.125 & 0.063\\
$\tu{OR}_{1,0\mid c}^{\tu{TE}}$     & 7.040  & 1.024 & 5.294 &  9.361& 0.000 & 7.121 &  1.045 &  5.341 &   9.494 & 0.000\\
\bottomrule
\end{tabularx}
\caption{Estimates, standard errors (SEs), 95\% confidence intervals (CIs) and p-values of the causal odds ratios for the mediation scheme of Figure~\ref{fig:data}}\label{tab:scen2}
\end{table}

As a sensitivity analysis, we have also recomputed the effect estimates after widening the set of observed confounders with some other potentially relevant variables excluded in the first place like income, value of family assets, gender and marital status. Also these results are in line with the values in Table~\ref{tab:scen2}. Though this tends to confirm the validity of our causal estimates, like in every empirical study the absence of unobserved confounding cannot be guaranteed with certainty, and results have to be interpreted with caution.

%% file: concl.tex
\section{Conclusions}\label{sec:concl}
We have focussed on causal mediation for binary outcomes, deriving novel parametric expressions for natural direct and indirect effects, on the odds ratio scale, for settings where the mediator is a binary random variable. The proposed formulae are exact in the sense that they do not require the rare outcome assumption and are suitable for the two different natural effect decompositions existing in the casual mediation literature. Furthermore, they are written in order to distinctly maintain the link between natural effects and their pathway-correspondent coefficients of the logistic regression models assumed to govern the data generating process. In particular, in line with~\cite{VdWVan2010} and~\cite{ValeriVdW2013}, we have isolated the controlled direct effect as part of the natural direct effect. We have also formalized the expressions of the approximate standard errors of the causal effect estimators, obtained via the delta method. The formulae for both the causal effects and the standard errors generalize the existing ones, especially with regard to the presence of parametric interactions between the exposure (and the mediator) and the confounders in the logistic models. 

As an illustration, we have applied the derived formulae to a dataset coming from a microcredit experiment performed in Bosnia and Herzegovina~\citep{augsburg2015impacts}, where it is plausible to think that the effect of randomly allocated microloans on client's bankability (i.e., the capability to obtain loans from financial institutions) at a 14-month follow up might be mediated by whether or not the client owns an active business. The causal effects estimated for the microcredit data are conditional on client's age, educational level and number of active loans at baseline. These are the mediator-outcome confounder variables we have chosen by combining empirical analyses and subject matter knowledge. The results we have obtained are rather stable across the patterns of these variables. Specifically, all the total causal effect odds ratios lie around 7.10, but only a small (and barely significant) part of such total effects appears to be mediated by business ownership.

The counterfactual approach to mediation analysis requires strong assumptions about confounding which might be difficult to meet in practice. Therefore, it would be interesting to adapt to the present context the existing methods of sensitivity analysis with respect to unobserved confounding. In particular, the interval identification method introduced by~\cite{Lindmarketal2018} for probit regression, based on the proposal of~\cite{Genbacketal2014}, could be adapted to logistic models. Other promising extensions involve the case of multiple mediators, for which associational decompositions have been recently derived.

%% file: appendix.tex
\section{Mathematical derivation of $\tu{OR}^{\tu{PNDE}}_{x,x^{\star}}$ and $\tu{OR}^{\tu{TNIE}}_{x,x^{\star}}$}\label{app:math}
Given the standard causal inference assumptions of Section~\ref{sec:dec}, the natural effects can be non-parametrically identified by using Pearl's mediation formula~\citep{Pearl2001,Pearl2010med}. For a binary mediator, the expression identifying the pure natural direct effect is
{\small
\[
\tu{OR}_{x,x^{\star}}^{\tu{PNDE}} = \frac{\overbrace{\sum_{w}P(Y=1\mid X=x,W=w)P(W=w\mid X=x^{\star})/\sum_{w}P(Y=0\mid X=x,W=w)P(W=w\mid X=x^{\star})}^{Q_1}}{\underbrace{\sum_{w}P(Y=1\mid X=x^{\star},W=w)P(W=w\mid X=x^{\star})/\sum_{w}P(Y=0\mid X=x^{\star},W=w)P(W=w\mid X=x^{\star})}_{Q_2}}
\]
}
Given the parametric models assumed, the numerator of the expression above can be written as
{\footnotesize
\[
\begin{split}
Q_1 &= \frac{P(Y=1\mid X=x,W=1)P(W=1\mid X=x^{\star})+P(Y=1\mid X=x,W=0)P(W=0\mid X=x^{\star})}{P(Y=0\mid X=x,W=1)P(W=1\mid X=x^{\star})+P(Y=0\mid X=x,W=0)P(W=0\mid X=x^{\star})} \\
   &= \frac{\frac{\exp\{\beta_0+\beta_w+(\beta_x+\beta_{xw})x\}}{1+\exp\{\beta_0+\beta_w+(\beta_x+\beta_{xw})x\}}\times\frac{\exp(\gamma_0+\gamma_xx^{\star})}{1+\exp(\gamma_0+\gamma_xx^{\star})}+\frac{\exp(\beta_0+\beta_xx)}{1+\exp(\beta_0+\beta_xx)}\times\frac{1}{1+\exp(\gamma_0+\gamma_xx^{\star})}}{\frac{1}{1+\exp\{\beta_0+\beta_w+(\beta_x+\beta_{xw})x\}}\times\frac{\exp(\gamma_0+\gamma_xx^{\star})}{1+\exp(\gamma_0+\gamma_xx^{\star})}+\frac{1}{1+\exp(\beta_0+\beta_xx)}\times\frac{1}{1+\exp(\gamma_0+\gamma_xx^{\star})}} \\
   &= \frac{\exp\{\beta_0+\beta_w+(\beta_x+\beta_{wx})x\}\exp(\gamma_0+\gamma_xx^{\star})\{1+\exp(\beta_0+\beta_xx)\}+\exp(\beta_0+\beta_xx)[1+\exp\{\beta_0+\beta_w+(\beta_x+\beta_{xw})x \}]}{\exp(\gamma_0+\gamma_xx^{\star})\{1+\exp(\beta_0+\beta_xx)\}+1+\exp\{\beta_0+\beta_w+(\beta_x+\beta_{xw})x \}} \\
   &= \exp(\beta_0+\beta_xx)A_{x,x^{\star}}.
\end{split}
\]
}
For the denominator, an analogous calculation leads to $Q_2 = \exp(\beta_0+\beta_xx^{\star})A_{x^{\star},x^{\star}}$ and therefore to $\log\tu{OR}_{x,x^{\star}}^{\tu{PNDE}}=\log Q_1 - \log Q_2 = \beta_x(x-x^{\star}) + \log(A_{x,x^{\star}}/A_{x^{\star},x^{\star}})$, which proves Equation~\eqref{eq:ndenocov}. Derivations for the total natural indirect effect are similar since we have
{\small
\[
\tu{OR}_{x,x^{\star}}^{\tu{TNIE}} = \frac{\overbrace{\sum_{w}P(Y=1\mid X=x,W=w)P(W=w\mid X=x)/\sum_{w}P(Y=0\mid X=x,W=w)P(W=w\mid X=x)}^{Q_3}}{\underbrace{\sum_{w}P(Y=1\mid X=x,W=w)P(W=w\mid X=x^{\star})/\sum_{w}P(Y=0\mid X=x,W=w)P(W=w\mid X=x^{\star})}_{Q_1}},
\]
}
with $Q_3=\exp(\beta_0+\beta_xx)A_{x,x}$, leading to $\log\tu{OR}_{x,x^{\star}}^{\tu{TNIE}}=\log Q_3 - \log Q_1 = \log(A_{x,x}/A_{x,x^{\star}})$, that is, Equation~\eqref{eq:nienocov}.

To prove that $A_{x,x}$ equals the inverse risk ratio term in~\eqref{eq:invrr}, let
\[
\begin{split}
g_y(x)= y(\beta_{w}+\beta_{xw}x)+ \log \left( \frac{1+\exp(\beta_{0}+\beta_{x}x)}{1+\exp(\beta_{0}+\beta_{x}x+\beta_{w}+\beta_{xw}x)} \right)+\gamma_{0}+\gamma_{x}x
\end{split}
\]
be the parametric expression for $\log \frac{P(W=1\mid Y=y, X=x)}{P(W=0\mid Y=y, X=x)}$ given by~\cite{StangheDoro2018}. Then, it is straightforward to prove that
\[
\begin{split}
A_{x,x} = \frac{1+\exp\{g_1(x)\}}{1+\exp\{g_0(x)\}},
\end{split}
\]
with the right-hand side term being indeed the probability ratio in~\eqref{eq:invrr}.

The algebraic developments above remain unchanged once confounding-removing covariates $C$ are added, provided that linear predictors are suitably modified. Specifically, the conditional versions of $Q_1$, $Q_2$ and $Q_3$ become
\[
\begin{split}
Q_{1\mid c} &= e_y(x,0,z)A_{x,x^{\star}\mid c} \\
Q_{2\mid c} &= e_y(x^{\star},0,z)A_{x^{\star},x^{\star}\mid c} \\
Q_{3\mid c} &= e_y(x,0,z)A_{x,x\mid c},
\end{split}
\]
where $e_y(x,w,z)$ and $A_{x,x^{\star}\mid c}$ are as in Section~\ref{sec:dec}. The derivation of Equations~\eqref{eq:ndecov} and~\eqref{eq:niecov} is then immediate. Notice that this approach can be immediately generalized to account for multiple confounders; it suffices to replace $z$ with $\bm{z}=(z_{1},\dots,z_{p})'$ and $v$ with $\bm{v}=(v_1,\dots,v_q)'$ in the formulas above, substituting every product involving $z$ and $v$ with the corresponding row-column product (for instance, $\beta_{z}z$ is replaced by $\bm{\beta}_{z}'\bm{z}$ with $\bm{\beta}_{z}=(\beta_{z_{1}},\dots,\beta_{z_{p}})'$ and so on).


\section{Natural effects and the rare outcome assumption}\label{app:limits}
Recalling that $e_y(x,w)=\exp(\beta_0+\beta_xx+\beta_ww+\beta_{xw}xw)$ and $e_w(x)=\exp(\gamma_0+\gamma_xx)$, the odds ratio pure natural direct effect is given by
\[
\begin{split}
\tu{OR}^{\tu{PNDE}}_{x,x^{\star}} &= \exp\{\beta_x(x-x^{\star})\} \\
&\times \frac{e(\beta_w+\beta_{xw}x)e_w(x^{\star})\{1+e_y(x,0)\}+1+e_y(x,1)}{e_w(x^{\star})\{1+e_y(x,0)\}+1+e_y(x,1)} \\
&\times \frac{e_w(x^{\star})\{1+e_y(x^{\star},0)\}+1+e_y(x^{\star},1)}{e(\beta_w+\beta_{xw}x^{\star})e_w(x^{\star})\{1+e_y(x^{\star},0)\}+1+e_y(x^{\star},1)}.
\end{split}
\]
If all the four terms $e_y(x,0)$, $e_y(x,1)$, $e_y(x^{\star},0)$ and $e_y(x^{\star},1)$ tend to zero, then the expression above reduces to
\begin{equation}\label{eq:pndelim}
\tu{OR}^{\tu{PNDE}}_{x,x^{\star}} \approx \exp\{\beta_x(x-x^{\star})\}\frac{1+e(\beta_w+\beta_{xw}x)e_w(x^{\star})}{1+e(\beta_w+\beta_{xw}x^{\star})e_w(x^{\star})},
\end{equation}
which is the expression in~\citet[p.150]{ValeriVdW2013}. The same logic applies to the odds ratio total natural direct effect, which is
\[
\begin{split}
\tu{OR}^{\tu{TNDE}}_{x,x^{\star}} &= \exp\{\beta_x(x-x^{\star})\} \\
&\times \frac{e(\beta_w+\beta_{xw}x)e_w(x)\{1+e_y(x,0)\}+1+e_y(x,1)}{e_w(x)\{1+e_y(x,0)\}+1+e_y(x,1)} \\
&\times \frac{e_w(x)\{1+e_y(x^{\star},0)\}+1+e_y(x^{\star},1)}{e(\beta_w+\beta_{xw}x^{\star})e_w(x)\{1+e_y(x^{\star},0)\}+1+e_y(x^{\star},1)}.
\end{split}
\]
Again, if all the four terms above tend to 0 this quantity tends to
\begin{equation}\label{eq:tndelim}
\tu{OR}^{\tu{TNDE}}_{x,x^{\star}} \approx\exp\{\beta_x(x-x^{\star})\}\frac{1+e(\beta_w+\beta_{xw}x)e_w(x)}{1+e(\beta_w+\beta_{xw}x^{\star})e_w(x)}
\end{equation}
in analogy with~\eqref{eq:pndelim}. The second factors of the right-hand sides of both~\eqref{eq:pndelim} and~\eqref{eq:tndelim} are equal to 1 if $\beta_{xw}=0$. This means that under the rare outcome assumption the odds ratio natural direct effects are given by a product between  $\tu{OR}^{\tu{CDE}(0)}_{x,x^{\star}}=\exp\{\beta_x(x-x^{\star})\}$ (governed by $\beta_x$ only) and a residual quantity governed by the interaction coefficient $\beta_{xw}$. This clear separation does not occur in the exact formulae, where both these residual terms are not generally equal to 1 when $\beta_{xw}=0$.

The odds ratio total natural indirect effect is
\[
\begin{split}
\tu{OR}^{\tu{TNIE}}_{x,x^{\star}} &= \frac{e(\beta_w+\beta_{xw}x)e_w(x)\{1+e_y(x,0)\}+1+e_y(x,1)}{e_w(x)\{1+e_y(x,0)\}+1+e_y(x,1)} \\
&\times \frac{e_w(x^{\star})\{1+e_y(x,0)\}+1+e_y(x,1)}{e(\beta_w+\beta_{xw}x)e_w(x^{\star})\{1+e_y(x,0)\}+1+e_y(x,1)}.
\end{split}
\]
Contrary to the case of natural direct effects, we notice that $e_y(x^{\star},0)$ and $e_y(x^{\star},1)$ are not present in the expression above. Thus, we only need $e_y(x,0)$ and $e_y(x,1)$ to tend to zero in order to obtain the expression in~\citet[p.150]{ValeriVdW2013}
\[
\tu{OR}^{\tu{TNIE}}_{x,x^{\star}} \approx \frac{\{1+e_w(x^{\star})\}\{1+e_w(x)e(\beta_w+\beta_{xw}x)\}}{\{1+e_w(x)\}\{1+e_w(x^{\star})e(\beta_w+\beta_{xw}x)\}}.
\]
This is particularly relevant when $X$ is binary, since it means that we only need $P(Y=1\mid X=1,W=0)\approx 0$ and $P(Y=1\mid X=1,W=1)\approx 0$, but necessarily $P(Y=1\mid X=0,W=0)\approx 0$ and $P(Y=1\mid X=0,W=1)\approx 0$.

Similarly, in the expression of the odds ratio pure natural indirect effect
\[
\begin{split}
\tu{OR}^{\tu{PNIE}}_{x,x^{\star}} &= \frac{e(\beta_w+\beta_{xw}x^{\star})e_w(x)\{1+e_y(x^{\star},0)\}+1+e_y(x^{\star},1)}{e_w(x)\{1+e_y(x^{\star},0)\}+1+e_y(x^{\star},1)} \\
&\times
\frac{e_w(x^{\star})\{1+e_y(x^{\star},0)\}+1+e_y(x^{\star},1)}{e(\beta_w+\beta_{xw}x^{\star})e_w(x^{\star})\{1+e_y(x^{\star},0)\}+1+e_y(x^{\star},1)}.
\end{split}
\]
the terms $e_y(x,0)$ and $e_y(x,1)$ are not present, so it suffices that $e_y(x^{\star},0)$ and $e_y(x^{\star},1)$ tend to zero to obtain
\[
\tu{OR}^{\tu{PNIE}}_{x,x^{\star}} \approx \frac{\{1+e_w(x^{\star})\}\{1+e_w(x)e(\beta_w+\beta_{xw}x^{\star})\}}{\{1+e_w(x)\}\{1+e_w(x^{\star})e(\beta_w+\beta_{xw}x^{\star})\}}.
\]
For the case of binary $X$, this means that the conditions $P(Y=1\mid X=0,W=0)\approx 0$ and $P(Y=1\mid X=0,W=1)\approx 0$ are needed, while $P(Y=1\mid X=1,W=0)\approx 0$ and $P(Y=1\mid X=1,W=1)\approx 0$ are not. Overall, it is possible to conclude that the two natural indirect effects need only different subsets of the rare outcome assumption (instead of the whole assumption as traditionally defined) to be identified by the approximate parametric formulae.


\section{Variance-covariance matrix of the estimated causal effects}\label{app:delta}
Denoting by $\bm{\beta}=(\beta_{0},\beta_{x},\beta_{z},\beta_{xz},\beta_{w},\beta_{xw},\beta_{wz},\beta_{xwz})'$ and $\bm{\gamma}=(\gamma_{0},\gamma_{x},\gamma_{v},\gamma_{xv})'$ the two vectors of model parameters, by $\bm{\Sigma}_{\hat{\bm{\beta}}}$ and $\bm{\Sigma}_{\hat{\bm{\gamma}}}$ the variance-covariance matrices of their estimators $\hat{\bm{\beta}}$ and $\hat{\bm{\gamma}}$, and by
\[
\bm{e}=(\tu{OR}^{\tu{PNDE}}_{x,x^{\star}\mid c},\tu{OR}^{\tu{TNIE}}_{x,x^{\star}\mid c},\tu{OR}^{\tu{TNDE}}_{x,x^{\star}\mid c},\tu{OR}^{\tu{PNIE}}_{x,x^{\star}\mid c},\tu{OR}^{\tu{TE}}_{x,x^{\star}\mid c})'
\]
the true causal effect vector, the first-order approximate variance-covariance matrix of the estimator
\[
\hat{\bm{e}}=(\hat{\tu{OR}}^{\tu{PNDE}}_{x,x^{\star}\mid c},\hat{\tu{OR}}^{\tu{TNIE}}_{x,x^{\star}\mid c},\hat{\tu{OR}}^{\tu{TNDE}}_{x,x^{\star}\mid c},\hat{\tu{OR}}^{\tu{PNIE}}_{x,x^{\star}\mid c},\hat{\tu{OR}}^{\tu{TE}}_{x,x^{\star}\mid c})'
\]
can be obtained as $V(\hat{\bm{e}}) = \bm{E}\bm{D}\bm{\Sigma}\bm{D}'\bm{E}'$, where $\bm{E}=\tu{diag}(\bm{e})$,
\[
\bm{\Sigma} = \begin{pmatrix} \bm{\Sigma}_{\hat{\bm{\beta}}} & \bm{0} \\ \bm{0} & \bm{\Sigma}_{\hat{\bm{\gamma}}} \end{pmatrix}
\]
and $\bm{D}$ is the matrix of derivatives $\bm{D} = \partial \log\bm{e}/\partial \bm{\theta}'$, with $\bm{\theta}=(\bm{\beta}',\bm{\gamma}')'$ denoting the whole parameter vector. To obtain $\bm{D}$, it is convenient to compute the row vector $\bm{d}_{x,x^{\star}\mid c}=\partial A_{x,x^{\star}\mid c}/\partial \bm{\theta}'$ first. To this end, it is worth to write $A_{x,x^{\star}\mid c}$ as
\[
A_{x,x^{\star}\mid c} = \frac{p_1p_2p_3 + p_4}{p_2p_3 + p_4},
\] 
with $p_1=\exp(\beta_w+\beta_{xw}x+\beta_{wz}z+\beta_{xwz}xz)$, $p_2=e_{w}(x^{\star},v)$, $p_3=1+e_{y}(x,0,z)$ and $p_4=1+e_{y}(x,1,z)$. Under this notation, the three key derivatives to compute are
\[
\begin{split}
d_{\beta_{0}}(x,x^{\star}\mid c) &= \frac{\partial A_{x,x^{\star}\mid c}}{\partial \beta_0} = \frac{\{p_1p_2(p_3-1)+p_4-1\}(p_2p_3+p_4)-(p_1p_2p_3 + p_4)\{p_2(p_3-1)+p_4-1\}}{(p_2p_3+p_4)^2} \\
d_{\beta_{w}}(x,x^{\star}\mid c) &= \frac{\partial A_{x,x^{\star}\mid c}}{\partial \beta_w} = \frac{(p_1p_2p_3+p_4-1)(p_2p_3+p_4) - (p_1p_2p_3 + p_4)(p_4-1)}{(p_2p_3+p_4)^2} \\
d_{\gamma_{0}}(x,x^{\star}\mid c) &= \frac{\partial A_{x,x^{\star}\mid c}}{\partial \gamma_0} = \frac{(p_1p_2p_3)(p_2p_3+p_4) - (p_1p_2p_3 + p_4)(p_2p_3)}{(p_2p_3+p_4)^2}, \\
\end{split}
\]
while the others can be written as functions thereof. Specifically, a compact form for $\bm{d}_{x,x^{\star}\mid c}$ is given by
\[
\bm{d}_{x,x^{\star}\mid c} = [(d_{\beta_{0}}(x,x^{\star}\mid c),d_{\beta_{w}}(x,x^{\star}\mid c))\otimes \bm{d}(x,z)\,,\,d_{\gamma_{0}}(x,x^{\star}\mid c)\bm{d}(x^{\star},v)],
\]
where $\otimes$ denotes the Kronecker product and, letting $\bm{I}_{2}$ be a diagonal matrix of order 2, $\bm{d}(a,b)$ is the row vector returned by the vector-matrix multiplication $\bm{d}(a,b) = (1\,,\,a)[(1\,,\,b)\otimes\bm{I}_{2}]$. The vectors $\bm{d}_{x,x\mid c}$, $\bm{d}_{x^{\star},x^{\star}\mid c}$ and  $\bm{d}_{x^{\star},x\mid c}$ can be calculated applying the same formulas above to $A_{x,x\mid c}$,  $A_{x^{\star},x^{\star}\mid c}$ and $A_{x^{\star},x\mid c}$ respectively. Then, the matrix $\bm{D}$ can be obtained as
\[
\bm{D}=\begin{pmatrix} \bm{d}_1 + \bm{d}_2 \\ \bm{d}_3 \\ \bm{d}_1+\bm{d}_4 \\ \bm{d}_5 \\ \bm{d}_6 \end{pmatrix},
\]
where
\[
\begin{split}
\bm{d}_2 &= \bm{d}_{x,x^{\star}}/A_{x,x^{\star}} - \bm{d}_{x^{\star},x^{\star}}/A_{x^{\star},x^{\star}}\\
\bm{d}_3 &= \bm{d}_{x,x}/A_{x,x} - \bm{d}_{x,x^{\star}}/A_{x,x^{\star}}\\
\bm{d}_4 &= \bm{d}_{x,x}/A_{x,x} - \bm{d}_{x^{\star},x}/A_{x^{\star},x}\\
\bm{d}_5 &= \bm{d}_{x^{\star},x}/A_{x^{\star},x} - \bm{d}_{x^{\star},x^{\star}}/A_{x^{\star},x^{\star}}\\
\bm{d}_6 &= \bm{d}_{x,x}/A_{x,x} - \bm{d}_{x^{\star},x^{\star}}/A_{x^{\star},x^{\star}}\\
\end{split}
\]
while $\bm{d}_1$ is a row vector of the same length of $\bm{\theta}$ with all its components set to zero but the ones in the positions of $\beta_{x}$ and $\beta_{xz}$, worth $x-x^{\star}$ and $z(x-x^{\star})$ respectively. Again, extension to multiple confounders is immediate provided that $\bm{\beta}$ and $\bm{\gamma}$ are extended as follows:
\[
\begin{split}
\bm{\beta} &= (\beta_{0},\beta_{x},\beta_{z_{1}},\dots,\beta_{z_{p}},\beta_{xz_{1}},\dots,\beta_{xz_{p}},\beta_{w},\beta_{xw},\beta_{wz_{1}},\dots,\beta_{wz_{p}},\beta_{xwz_{1}},\dots,\beta_{xwz_{p}})'\\ 
\bm{\gamma} &= (\gamma_{0},\gamma_{x},\gamma_{v_{1}},\dots,\gamma_{v_{q}},\gamma_{xv_{1}},\dots,\gamma_{xv_{p}})'.
\end{split}
\]
Clearly, in finite-sample analyses one has plug in the estimates $\hat{\bm{\beta}}$ and $\hat{\bm{\gamma}}$ everywhere in the formulae above to obtain the estimated variance/covariance matrix $\hat{V}(\hat{\bm{e}})$.